# Situational Awareness in Indian Power Grid using Synchrophasor Data

Makarand Sudhakar Ballal, Dept. of Electrical Engineering, VNIT, Nagpur

*Abstract*— **Wide Area Measurement Systems (WAMS) can guide system operators' to increase their situational awareness by expanding observability of their supervise area and adjoining systems. Power system oscillations in the electrical grid are a matter of contention. It may transform the stable operation of the power system into an unstable system when oscillations grow and if no corrective actions are taken to mitigate them. This paper manifests the utilization of the synchrophasor technology for situational awareness by oscillation detection. An algorithm is proposed for the low-frequency oscillation-related analysis based on Prony and Fast Fourier Transform (FFT) methods. It gives better insight during the grid disturbance and other events related to power system operation. The effectiveness of this algorithm is validated by two case studies in the Western Regional Indian power grid.**

## 1. Introduction

**W**hen a stone is thrown in a quiet pond, a circular wave emerges on the surface of the water. It gently moves across the pond in a concentric pattern (see Fig. 1). Physicists call these ripples "surface waves". The wave produces an up and down squabble of the water on its surface. The time taken for one complete up and down motion is termed as the wave's period *T*. Thus, the wave's frequency is $f = 1/T$. These surface waves are known by oscillations as they oscillate at certain frequencies. The oscillation frequencies at points A, B, and C are different. It is because the wave propagates at different time intervals [1]. Similar to this, whenever a disturbance occurs in a power system, it also produces oscillations. In an electrical power system, alternators and loads are interconnected through a transmission and distribution network. The alternators are synchronized and operated at a defined band of frequency. Any disturbance such as faults or transients in the power system can change the speed of anyone alternator from the synchronous speed. This affects the change in power of all other AC generators attached to the system. Under such contingencies, the system sustains synchronous speed by employing better control action. This includes appropriate control actions of the controllers in the exciter or governing system. However, if the controllers' actions are insufficient, this gives rise to low-frequency oscillations (LFOs). Especially, the high-speed field excitation system turns to cripples the damping of LFOs. In a power system, if LFOs continue for a relatively long period, these threaten the stability of the system. In this era, it is possible to measure these low-frequency oscillations by synchrophasors also known as phasor measurement units (PMUs) for the applications of Wide Area Measurement Systems (WAMS). The assertive oscillation modes signify in a frequency range of 0.1 to 2.0 Hz. Several local generators oscillate in the local mode and many generators oscillate between areas in the inter-area mode [2].

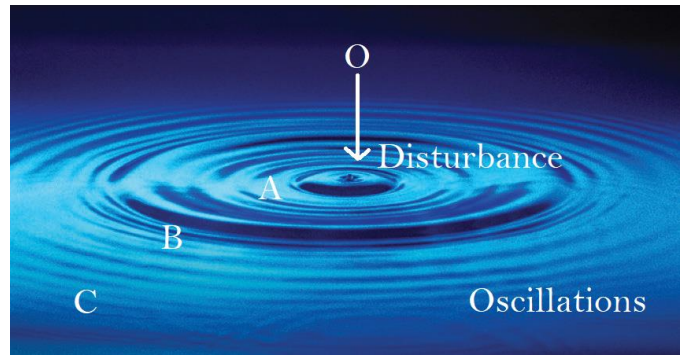

Fig.1: Disturbance and generation of oscillations in standstill water.

WAMS are making their way into power utility systems impelled by engrossment for pliable decisions. They are required for bettering system trade, and its reliability. Synchrophasor data from WAMS available at adequately high quantity shows the real state of the power system. It offers an exquisite solution to resolve the practical challenges of contemporary power systems.

WAMS is used for situational awareness by improving the observability of the defined control area and adjoining systems. The changing situations of the defined control area can be examined expeditiously by combining the geographical system data. The purpose of situation awareness is to comprehend the practical information, to appraise the susceptibility of the extant state, and to determine the issue for corrective actions, if imperative. The situation awareness comprises three stages such as attention, apprehension, and prolongation. Statistical techniques are applied to find the recurrent increased levels of phase angle deviations from synchrophasor data is explained in [3]. This permits real-time record, determination of exemplary conditions, and prediction of security margins. The methodology reported in [4] is an event-aligned adaptable moving window approach used to promote situational awareness. It accomplishes a curve fitting technique using a weighted exponential function. It is based on variable sliding window housing diverse event types. In [5], the developments of the frequency monitoring network; sensor, and server design are reviewed. Most of these applications are accommodated in the existing power utility services to boost up the situational awareness expertise for efficient grid control operations. In [6], a double layer dynamic optimal PMU measurement algorithm is explained. It is established on auxiliary voltage control as well as provincial smart grid measurement. It is applied to detect auxiliary voltage control areas, respective buses for measurement, and the placement of PMUs. An approach for the detection and sensual localization of numerous crises using PMU data is manifested in [7]. It is based on a Teager-Kaiser energy method. For event allocation, a time series-dependent technique using energy similarity measure (ESM) is applied in [8]. Islanding condition has been determined to realize situational awareness. This technique uses principal component analysis (PCA) for nonlinearities among variables and for data condensation. The Multi-resolution Teager energy method is applied to detect islanding conditions.

A scalable synchrophasor network using wireless LAN technology is explained in [9]. It is designed to implement network performance for a practical system. It reinforces integrated controllers for situational awareness in the transmission network when connected with distributed generations. In [10], a double-step abnormality detection technique that works out raw PMU data by the Map-Reduce paradigm is described. It was executed on a multicore system to process a dataset received from practical PMUs. It exhibits the appropriateness of designing irregularity identification algorithms for situational awareness. The practical event location is of crucial importance for situational awareness, stability studies, and operational control of the power grid. An adaptive online event location approach is explained in [11]. The anisotropy of the frequency propagation speed (FPS) is considered. This approach geographically segregates the power system into various areas. Thereafter, it determines the FPS vector for events appearing in each area. A conscientious identification of faults on the transmission system, using Phaselets computed from PMU measurements is given in [12]. It is made with visual displays used by operators for improving situational awareness to take control and protection decisions. Synchronous machine's dynamic state estimation determined by synchrophasor data using the extended Kalman filter (EKF) approach is explained in [13]. It figures out the angle of the machines at variables speed that presents a correct view of the power system for situational awareness. A framework reported in [14] is for the improvement of grid operations by acquiring situational awareness and protection due to WAMS. It works on the analysis of prominent disturbance signatures determined from PMU data.

Observability of the grid is achieved by certain routine and self-supervision tests at predefined intervals are described in [15]. Situational awareness can be possible by an intelligent maintenance scheduling algorithm. It is useful when the synchrophasors are under maintenance. System to enhance operator's situational awareness given in [16] determines the voltage stability margin from PMUs data. Cubic spline extrapolation technique with Thevenin equivalent model is adapted. To detect oscillations at early stages, in [17] a cross-coherence technique using multiple-channel PMU data is elaborated. By visual inspection the oscillations in coherence spectra are determined. Thresholds shall be updated for better situational awareness. An approach for the identification and classification of multiple events in practical power systems such as islanding, dislocation of load, and backing down of generation, etc. is reported in [18]. It uses principal component analysis and a sliding window technique to assimilate the time-varying nature of power systems. The optimal placement of PMUs in the power network is determined by a binary genetic algorithm is given in [19]. It uses multistage installation and the performance is dependent on the cost set by the utility. The PMU data received from prime important substations underwent data compression by singular value and the eigenvalue decompositions are explained in [20]. Data of two major events in the Maharashtra State Electricity Transmission Company Limited (MSETCL) transmission grid is considered in this study. How big data used for making the right decision at right time to improve situational awareness is given in [21]. It elaborates the correlation among the synchrophasor data collected from eight important 400 kV substations in the MSETCL grid. The operational experiences of three important System Integrity and Protection Schemes (SIPS) in India are presented in [22-24]. The shortcomings of these SIPS are detected. An algorithm is employed to resolve these discrepancies by the synchrophasor technology. This paper presents an application algorithm used to alert the system's operator and thereby it helps to improve situational awareness. The extensive contribution of this research article is specified below.

    i. Prony and FFT methods are smartly used in the development of this algorithm.
    ii. It only requires PMU data for the detection of various modes of oscillations.
    iii. There is no necessity for any past data or projection and or estimation approach.
    iv. Two practical cases are demonstrated to show the effectiveness of this algorithm.

The rest of the paper is arranged as follows. The development of WAMS in India and the system under study is explained in Section 2. The research motivation, application of Prony and FFT methods, and the proposed algorithm for situational awareness are discussed in Section 3. In Section 4, the potency of the proposed application algorithm is revealed by considering two case studies. The end section 5 outlines the concluding remarks.

## 2. WAMS in India and System under Study

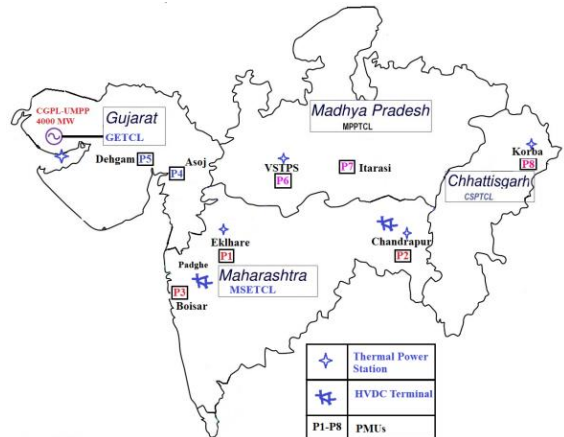

Fig.2: Major PMU locations in WR grid in India [25].

Synchrophasors are placed in India as part of a Union Government Unified Real-Time Dynamic State Measurement (URTDSM) project. The project has been assimilated to accomplish applications of PMUs in the Indian power grid operation. This national WAMS project covers the fixing of 1740 PMUs in ultra and extra-high voltage substations and major generating power plants across India. At State, Regional and National level, the 32 Phasor Data Concentrators (PDC) has to be equipped. For better management, the Indian national power grid is topographically segregated into five regional grids. These regional grids are Northern, Eastern, Western, North Eastern, and Southern grids. Therefore, the five super PDCs have been placed at the respective five regional load dispatch centers (RLDCs). At the end of March 2021, PMU commissioned is approximately 1500 in India [25]. The Western Regional power grid of India is dispersed in the states of Chhattisgarh, Goa, Gujarat, Madhya Pradesh, Maharashtra, and union territories of Div, Daman, Dadra, and Nagar Haveli.In addition to the Central Transmission Utility (CTU) of India, i.e. Power Grid Corporation of India Limited (PGCIL), leading transmission service providers operating in WR grid are Gujarat Electricity Transmission Company Limited (GETCL) in Gujarat, Chhattisgarh State Power Transmission Company Limited(CSPTCL) in the Chhattisgarh, Madhya Pradesh Power Transmission Company Limited(MPPTCL) in the Madhya Pradesh and Maharashtra State Electricity Transmission Company Limited (MSETCL) in Maharashtra. Fig.2 shows geographical viable boundaries for main state transmission companies in the WR of India. The position of prime PMUs implied as P1…P8 from the WR grid are shown and their data is applied for the analysis purpose. The substations considered for case studies in this paper are also depicted in Fig. 2. They are a super thermal power plant in Gujarat operated by Coastal Gujarat Power Limited (CGPL) under the Ultra Mega Power Project(UMPP) at Mundra, and ± 500 kV HVDC terminals at Padghe in Maharashtra [25].

GETCL installed ten PMUs at 400 kV and thirteen PMUs at 220 kV major substations. It covers the maximum transmission network of the State. MSETCL has also PMUs at twelve 400kV and three 220kV level substations in its grid. Placement of PMU in 15 EHV substations covers 67% of the MSETCL network [20]. The State Load Despatch Centre (SLDC) of Gujarat and Maharashtra are located at Vadodara and Kalwa (Mumbai) respectively. They have Phasor Data Concentrator (PDC), visional, and data storage software uploaded under WAMS. The real-time data transfer from PMUs for parameters like line voltages ($V$), currents ($I$), active power ($P$), reactive power ($Q$), frequency ($f$), rate of change of frequency ($df/dt$), bus voltage angle delta ($\delta$), and status of circuit breakers is done from PMUs located at different important substations in the state grid to respective PDC at SLDC. For the analysis of situational awareness, this data is required. This data is processed through a specialized algorithm. The following section describes the techniques used in this paper to process this data. It describes the development of an algorithm used to judge the situational awareness by estimating low-frequency modes of oscillations.

## 3. Situational Awareness by Oscillation Analysis

The design and development of a real-time power system oscillation monitoring algorithm is an important part of situational awareness studies using WAMS technologies. This section describes the research motivation, methodology selection, and application algorithm for situational awareness.

*3.1 Research Motivation*

It is noticed that the smart grid comprises distinct state-of-the-art technologies allowing impressive protocol for renewable energy submission, variable critical and non-critical loads, and energy storage. This notion overtures the prudent and competent operation of imminent power systems including generation, transmission, distribution, and consumption. Efficacious grid administration is possible if the appropriate and fruitful data is available by the system operator. This data contains the amplitude and angle of the current and voltage phasors, frequency, rate of change of frequency (ROCOF), apparent load flow, the position of the circuit breakers, oscillations during normal and abnormal conditions, etc. This data is used to investigate power system dynamics. Its applications cover wide-area monitoring, detection of fault location, detection of islanding, state estimation, operation of protective relaying, etc. These applications are useful in situational awareness as they provide more insight into the power system.

To implement smart grid technology, the Indian power grid has also taken major steps. It has been executed after blackouts in the northern grid and north-eastern grid in July-2012 by launching a pilot project of WAMS applications. State and central power transmission companies have connected PMUs for wide-area monitoring, protection, and control (WAMPAC). Unified Real-Time Dynamic State Measurement (URTDSM) project is the Indian WAMS project that commenced to install 1740 PMUs in the ultra and extra-high voltage (EHV) substations. Thus, the Department of Energy (DOE) is giving considerable attention to securing the power grid. PMUs are connected at many prime substations in the WR under the URTDSM project begun to provide valuable data. The data received from PMUs, motivate the author to develop an application algorithm for situational awareness.

*3.2 Methodology Selection*

In several Smart Grid ventures, some distinguished techniques were adopted for the detection of power system oscillation. These techniques are interpreted and compared to diagnose the best in terms of (i) accuracy in the recognition of the oscillation's mode, (ii) modesty in the attainment, (iii) strength to present visual proclivity, (iv) capacity to find the most related state variable to every oscillation's mode, (v) immune to the noise, and (vi) self-checking capability. A literature survey revealed that the Prony and FFT methods can be an acceptable choice for the prospective algorithm.

1. Prony Method

The Prony method executes in both forward and backward style. The final modes are anointed from the assortment by such a process. This process facilitates the Prony method to allocate incidental modes and eliminate the noise effects. Prony is intrinsically a self-verifiable method as it determines the modes of a signal and then reassembles the vital signal from the anticipated mode. Comparing the original and the rejuvenated signal is a commodious way to justify its performance. These above demeanors make the Prony method an efficient one for field data applications. However, it does not give viewable intuitiveness on the contribution of numerous system modes in every oscillation mode. This is imperative because power system operators are interested in viewable information instead of quantitating data [26-27].

Let us consider $y(t)$ to be a signal having $N$ evenly-spaced samples [28]. It can be fit in terms of Prony's method as

$$y(t) = \sum_{n=1}^{N} A_n e^{\sigma_n t} \cos(2\pi f_n t + \theta_n) \tag{1}$$

Where $A_n$ indicates the magnitude, $\sigma_n$ indicates the damping factor, $f_n$ represents frequency, and $\theta_n$ shows the phase angle of a signal. These four components can be determined from the state space depiction of an evenly sampled signal record. Between each sample, let $T$ is the time interval. Prony's method gives the dissolution of a signal with $M$ complex exponentials. Applying Euler's theorem and assign $t = MT$, the samples of $y(t)$ are expressed as

$$y_M = \sum_{n=1}^{N} B_n \lambda_n^M \tag{2}$$

$$B_n = \frac{1}{2} A_n e^{j\theta_n} \tag{3}$$

$$\lambda_n = e^{(\sigma_n + j2\pi f_n)T} \tag{4}$$

Prony analysis consists of three steps.

(i) In the introductory step, the coefficients of a linear prediction model (LPM) are computed. The LPM of order $N$, given in (5), is assembled to fit the coequally sampled $y(t)$ with length $M$. Generally, the $M$ should be three times more than the order $N$:

$$y_m = a_1 y_{M-1} + a_2 y_{M-2} + \ldots + a_N y_{M-N} \tag{5}$$

(ii) In the middle step, roots ($\lambda_n$) of the characteristic polynomial as given in (6) are obtained. These roots are related to LPM determined in the first step. By using (4), the damping factor ($\sigma_n$) and frequency ($f_n$) are obtained.

$$\lambda^N - a_1 \lambda^{N-1} - \ldots - a_{N-1} \lambda_{-a_N}$$
$$= (\lambda - \lambda_1)(\lambda - \lambda_2) \ldots (\lambda - \lambda_n) \ldots (\lambda - \lambda_N) \tag{6}$$

The damping factor ($\sigma_n$) is

$$\sigma_n = \frac{\mathrm{Re}(\log \lambda_n)}{T} \tag{7}$$

The frequency is

$$f_p = \frac{\mathrm{Im}(\log \lambda_n)}{2\pi T} \tag{8}$$

(iii) In the third and last step, a least square method is adopted for solving the magnitudes and the phase angles of the signal. The solved roots of $\lambda_n$ are used to get (9) and by using (2).

$$Y = \phi B, \tag{9}$$

$$\mathbf{Y} = [y_0 \ y_1 \ \ldots \ y_{M-1}]^T, \tag{10}$$

$$\phi = \begin{bmatrix} 1 & 1 & \ldots & 1 \\ \lambda_1 & \lambda_2 & \ldots & \lambda_N \\ \vdots & \vdots & \ldots & \vdots \\ \lambda_1^{M-1} & \lambda_2^{M-1} & \ldots & \lambda_N^{M-1} \end{bmatrix}, \tag{11}$$

$$\mathbf{B} = [B_1 \ B_2 \ \ldots \ B_N]^T. \tag{12}$$

The amplitude ($A_n$) and phase ($\theta_n$) are determined from the variables $B_n$ as

$$A_n = \frac{2B_n}{e^{j\theta_n}} \tag{13}$$

$$\theta_n = \mathrm{Im}\left\{\log\left(\frac{2B_n}{A_n}\right)\right\} \tag{14}$$

The Prony analysis can figure out the damping factor is its main advantage. Thus it is possible to identify the transient harmonics very accurately.

2. FFT Method

The Fast Fourier Transform (FFT) is an analytical approach for revamping a time-domain digital signal toward a frequency-domain depiction of the relative size (amplitude) of various frequency suburbs in the signal. The FFT is a technique for doing this procedure very effectively. It may be determined using a comparably short fragment from a signal. The FFT is one of the utmost indispensable propositions in Digital Signal Processing [29].

Let us consider a power frequency signal $y(t)$ given as,

$$y(t) = A(\sin \omega t + \varphi) \tag{15}$$

If (15) is sampled $M$ times, then the sample set $\{y_m\}$ is given as

$$y_m = A \sin\left(\frac{2\pi m}{M} + \varphi\right) \tag{16}$$

Discrete Fourier Transform (DFT) is used in Fourier analysis. It is a distinct type of discrete transform. It transforms one function into another. This is known by the frequency-domain representation of the original time-domain function. A finite sequence of real or complex numbers is the input to the DFT. This makes the DFT useful for processing information stored in the memory of computers.

The FFT of sinusoid $x(t)$ given in (16) can be asserted in discrete form as

$$Y(k) = \frac{1}{M} \sum_{m=0}^{M-1} y(m) e^{-j2\pi m \frac{k}{M}} \tag{17}$$

In the frequency domain, $Y(k)$ has two parts, i.e. real and imaginary. In the polar form, the magnitude and phase angle of the signal is given as,

$$\mathrm{Mag}\, Y(k) = \sqrt{(\mathrm{Re}\, Y(k))^2 + (\mathrm{Im}\, Y(k))^2} \tag{18}$$

$$\mathrm{Phase}\, Y(k) = \tan^{-1} \frac{\mathrm{Im}\, X(k)}{\mathrm{Re}\, X(k)} \tag{19}$$

The Prony and FFT methods were applied to synchrophasor data for the detection of power system oscillations. Both the above techniques are used in the development of an application algorithm for the assessment of situational awareness. It is explained in the following sub-section.

*3.3 Application Algorithm*

The problem LPO in the 0.2–1.5 Hz range either due to system interdependence or heavy load conditions is always there in power systems. Instability may happen if these oscillations are not damped effectively and it may result in system destruction. Therefore, for dynamic system security, it is necessary to check the low-frequency modes and the damping coefficients of the raw signal.

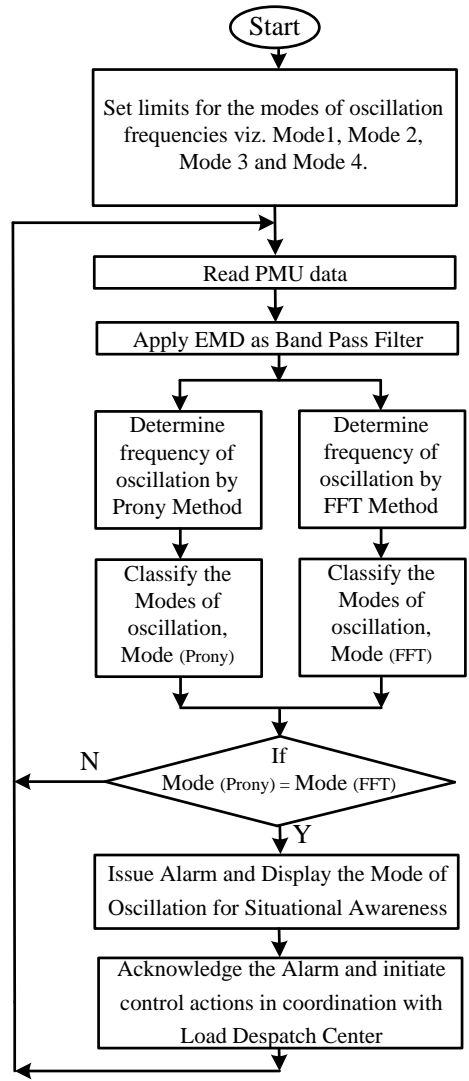

Fig. 3 Algorithm for the Situational Awareness

Low signal stability is the capability of synchronous machines of an integrated power system to preserve synchronism after being subjected to a small disturbance. This is classified as a phase angle associated instability problem. It depends on the capability to retain equilibrium between mechanical and electromagnetic torques of every synchronous machine connected to a power system. The fluctuation in electromagnetic torque of synchronous machines following a disturbance can be undertaken into two components. The first is a synchronizing torque component which is in phase with rotor angle deviation. Another is a damping torque component which is in phase with speed deviation. The non-oscillatory instability comes into existence because of insufficient synchronizing torque. And the insufficient damping torque draws low-frequency oscillations. LFOs are generally synchronous machine rotor angle oscillations having a frequency between 0.1 -2.0 Hz. They are categorized on the source of the oscillation. The main cause of electrical power oscillations is the unbalance between available power and power demand at a particular time interval. The power oscillations are non-observable in the olden days of power system development as the generators are tightly connected to loads. Nowadays, large demands of power to the furthest end of the system that brunt to transmit tremendous power through a long transmission line. These consequences increasing power oscillations. The phenomenon comprises mechanical oscillation of the rotor phase angle for a rotating frame. Change in the phase angle i.e. either increasing or decreasing with a low frequency will be resonated in power transferred from a generator. This is because the phase angle is paired with power transfer. The LFO can be classified as below [30].

1. Local modes of oscillations are correlated with the swinging of generating units at a power plant corresponding to the rest of the power system. These oscillations developed only to the limited part of the power system. Typically, their frequency range is 1-2 Hz. These oscillations in this frequency band are termed Mode 1 oscillations.

2. Interarea modes of oscillations are corresponding with the swinging of many machines in one part against machines in other parts of the system. It normally occurs in wobbly interconnected power systems by long tie lines. The frequency range is 0.1-1 Hz. These oscillations in this frequency band are termed Mode 2 oscillations.

3. Control or intra-plant modes of oscillations are associated with poorly tuned Power System Stabilizers (PSS). PSS finds use in interconnected networks in damping LFOs. The tuning study computes the optimum PSS settings, based on the particular synchronous machine, automatic voltage regulator settings, and system characteristics. These modes of oscillations have a higher frequency in the range of 1.5 to 8 Hz and are observable only within and nearby a generating plant. These oscillations in this frequency band are termed Mode 3 oscillations.

4. Torsional modes of oscillations are associated with the turbo-generator shaft. Torsional vibration is generally a concern in power transmission systems working rotating shafts or couplings where it can cause missteps if not controlled. These oscillations can impact the shaft torsional modes of the turbine generators [31]. The shaft system has three natural frequencies called torsional modes and they are generally more than 10 Hz. The oscillations more than 10 Hz are termed Mode 4 oscillations. However, the focus of this paper is more on the local, inter-area, and control or intra-plant oscillation modes that are measured during ambient conditions and event conditions when there is a large event like fault.

The frequency prediction based on Prony's method is highly sensitive to noise for field measurements when the signal-to-noise ratio is low. Thus the results are biased. Therefore, Empirical Mode Decomposition (EMD) is applied in this study. It is used as a band pass filter to remove high-frequency noise and the signal trend. EMD is a technique used for nonlinear, non-stationary signal processing [32]. It disintegrates the signal into a set of Intrinsic Mode Functions (IMFs). The constraints for an IMF are that its average value is zero. The number of peaks equals the number of zero crossings or uttermost differs by one. The step-by-step process to compute IMFs in a measured signal $y(t)$ is given below.

(i) Input the signal $y(t)$,
(ii) Determine the peaks,
(iii) Compute the upper ($e_{up}$) and lower ($e_{low}$) envelope of the signal,
(iv) Calculate the average value ($m(t)$) of the upper and lower envelope
(v) Determine the difference from the signals: $d(t) = y(t) - m(t)$
(vi) Repeat step (ii) – (v) with $d(t)$ as input, until it satisfies the conditions of an IMF
(vii) Fix $d(t)$ as an IMF, and deduct it from the input signal. Repeat the process with the residue as $r(t) = y(t) - d(t)$
(viii) Continue this process until there are no more peaks present in the signal.

The EMD output extracts modal components starting with the highest frequency and ending with the residual "trend" of the signal. By merging the IMFs with an average frequency in the oscillating frequency range from 0.1 to 2 Hz, the high-frequency components and the signal is excluded, and a band pass filter has been implemented.

The algorithm shown in Fig. 3 determines the modes of oscillations by Prony and FFT methods. The limits for the selection of modes of oscillations are set. The voltage signal received from PMU is filtered by the EMD Band Pass filter. This filtered signal is processed by Prony and FFT method simultaneously. The modes of frequency oscillations are identified and classified. If the mode of oscillation determined by both methods is matched, an alarm is issued for the information of the system operator for situational awareness. Thereafter, the operator shall initiate control actions in coordination with Load Dispatch Centre (LDC). The problem of LFO in the 0.2–1.5 Hz range either due to system integration or heavy load conditions is always there in power systems. Instability may happen if these oscillations are not damped effectively and it may result in system collapse. Therefore, for dynamic system security, it is necessary to check the low-frequency modes and the damping coefficients of the raw signal. The application of this algorithm is demonstrated in next section.

## 4. Results and Discussion

This section presents the applications of the above-described proposed situational awareness algorithm to the PMU data in the WR by following case studies. Two case studies are considered for this purpose. The first event is related to a blackout due to a disturbance in Coastal Gujarat Power Limited (CGPL)-Ultra Mega Power Project (UMPP) at Mundra. Another event is a disturbance in the HVDC system operation in the MSETCL grid.

### 4.1 Case Study-1: Blackout at CGPL-UMPP, Mundra

Ultra Mega Power Plant (UMPP) is installed at Mundra (Gujarat), and it is maintained by Coastal Gujarat Power Limited (CGPL). It consists of 5 generating units each of 800 MW, thus the total power generation is 4000 MW. It supplies power to two States in northern India viz. Punjab and Haryana and three States in western India viz. Rajasthan, Gujarat, and Maharashtra. A complete power shutdown due to an unstable power swing event occurred at CGPL on July 13, 2016 [25]. Fig. 4 illustrates the single line diagram (SLD) of this event with disturbances chronology in 400 kV substations.

Before this occurrence many important 400kV and 765kV transmission lines nearby CGPL-UMPP were out of service for maintenance. CGPL-UMPP is linked with the 400kV Varsana substation in the GETCL system. Affiliation of CGPL-UMPP with the rest of the system became weaker due to the bus fault at the 400kV Varsana substation. Tripping of all interconnections of CGPL-UMPP like that of 400kV CGPL-Bachau-1,400kV CGPL-Chornia, 400kV CGPL-Mansar, 400kV CGPL-Jetpur 1and 2, etc. resulted in the event. This further caused tripping of all CGPL-UMPP units nomenclature as, Unit-10 [760MW], Unit-30 [742MW], Unit-40[757MW], Unit-50[619MW] resulting in cumulative generation loss of 2878MW. This catastrophic failure tends to the blackout of the CGPL-UMPP and adjacent areas. The loss of generation resulted in frequency decay-causing $df/dt$ relay-based load shedding of 119MW in the GETCL system in Gujarat. The details of this occurrence are given in [25].

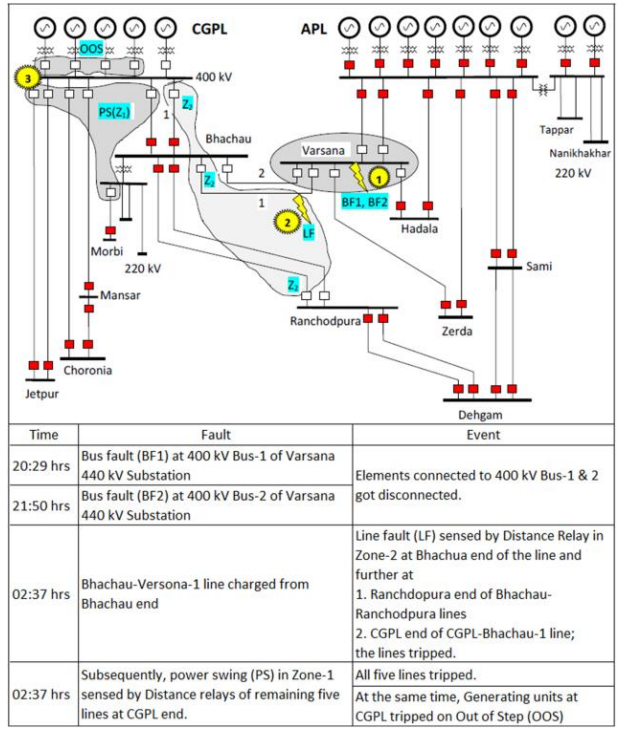

Fig 4. SLD with a history of a disturbance at CGPL-UMPP, Mundra [25].

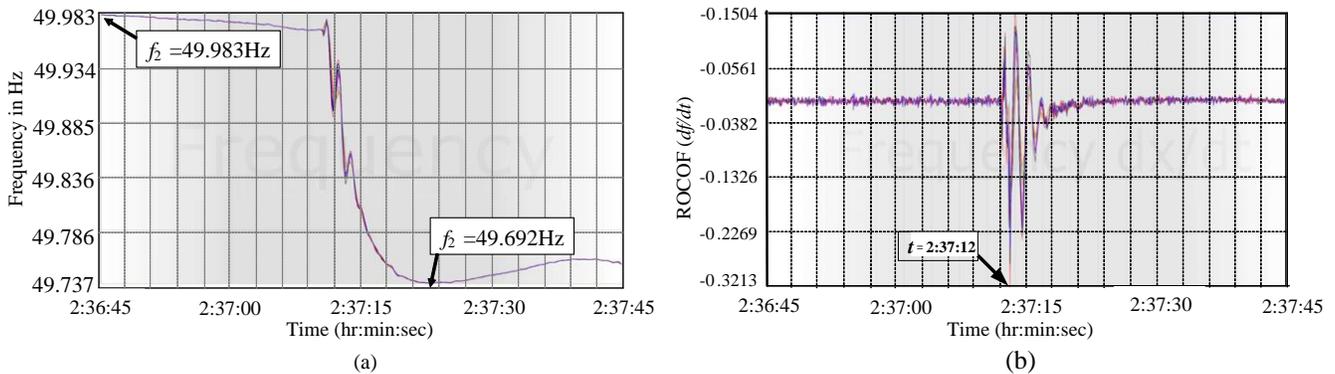

Fig 5 CGPL-UMPP occurrence indicated by PMUs at 220kV Boisar and Eklahre substations in MSETCL grid (a) Frequency behavior, and (b) ROCOF ($df/dt$).

Frequency behavior as sensed by PMUs in the MSETCL grid during CGPL-UMPP disturbance is shown in Fig. 5(a). The $df/dt$ as observed by PMUs at 220kV Boisar and 220kV Eklahre substations in the MSETCL grid during this event is depicted in Fig. 5(b). ROCOF ($df/dt$) as observed by PMUs at 220kV Boisar and 220kV Eklahre substations in the MSETCL system is -0.3213 Hz/s and -0.2790 Hz/s respectively. Frequency measured from PMUs across the MSETCL grid is used to carry out Prony and FFT analysis for understanding types of oscillatory modes excited during CGPL-UMPP occurrence. PMU data of duration 25s is considered for analysis purposes. Fig.6 shows 220kV Eklahre PMU data analysis plots in the MSETCL grid. Fig.6 (a) shows specific data range signal of frequency considered for Prony and FFT analysis. Fig. 6(b) shows a plot for Prony approximate and PMU measured frequency response. Fig.6(c) indicates a magnified part of the Fourier transform plot obtained for the 220kV Eklahre substation. Fig.7 shows 400kV Chandrapur substation PMU data analysis plots. Fig. 7(a) indicates Prony approximate and PMU measured frequency response. Fig.7 (b) shows the magnified part of the Fourier transform plot.

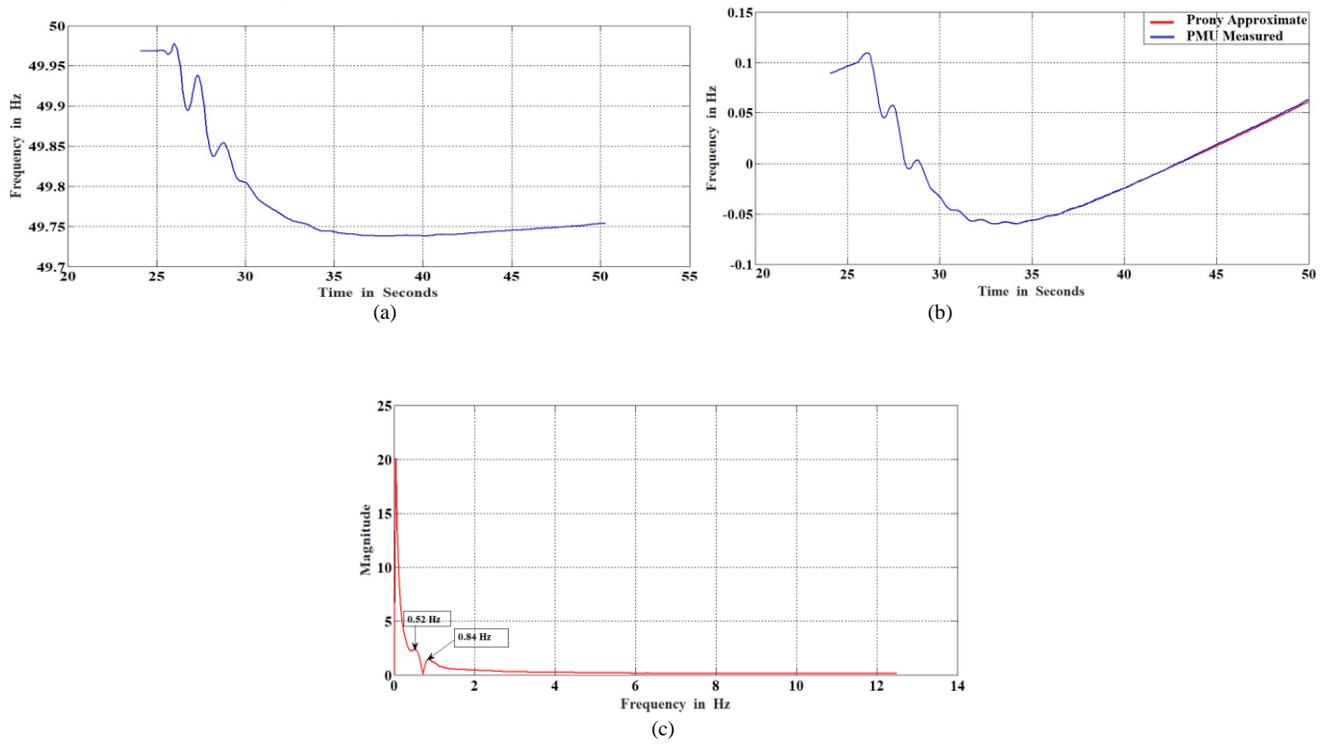

Fig.6. PMU data analysis plots for 220kV Eklahre substation (a) specific data range of PMU measured frequency considered for analysis, (b) plot for Prony approximate and PMU measured frequency response, and (c) FFT Plot indicating dominant frequency modes and related magnitude for frequency.

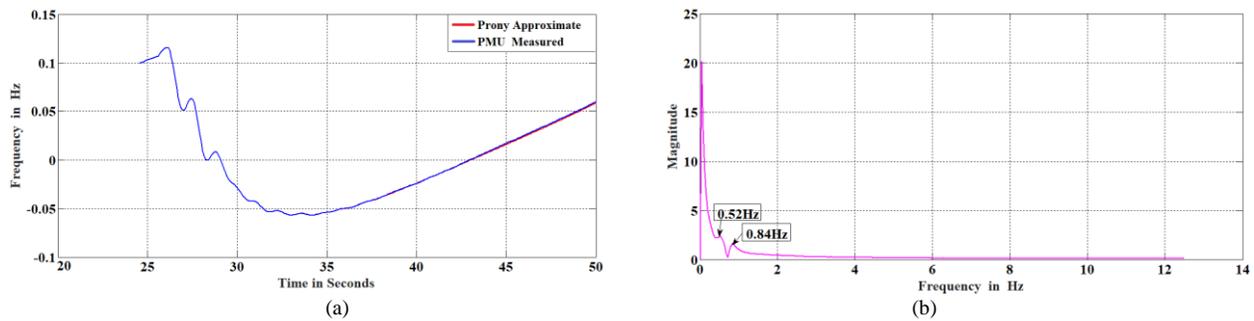

Fig.7. PMU data analysis plots for 400kV Chandrapur substation (a) specific data range of PMU measured frequency considered for analysis, (b) plot for Prony approximate and PMU measured frequency response, and (c) FFT Plot indicating dominant frequency modes and related magnitude for frequency.

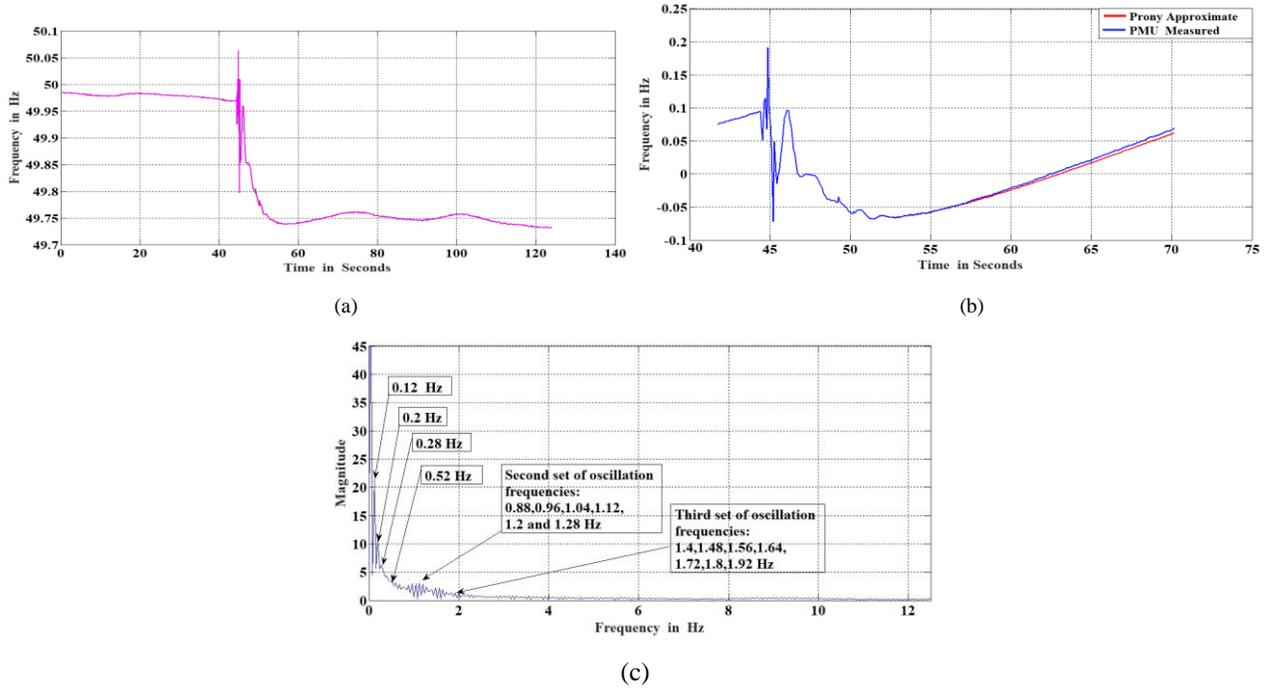

Fig.8. PMU data analysis plots for 400kV Asoj substation (a) specific data range of PMU measured frequency considered for analysis, (b) plot for Prony approximate and PMU measured frequency response, and (c) FFT Plot indicating dominant frequency modes and related magnitude for frequency.

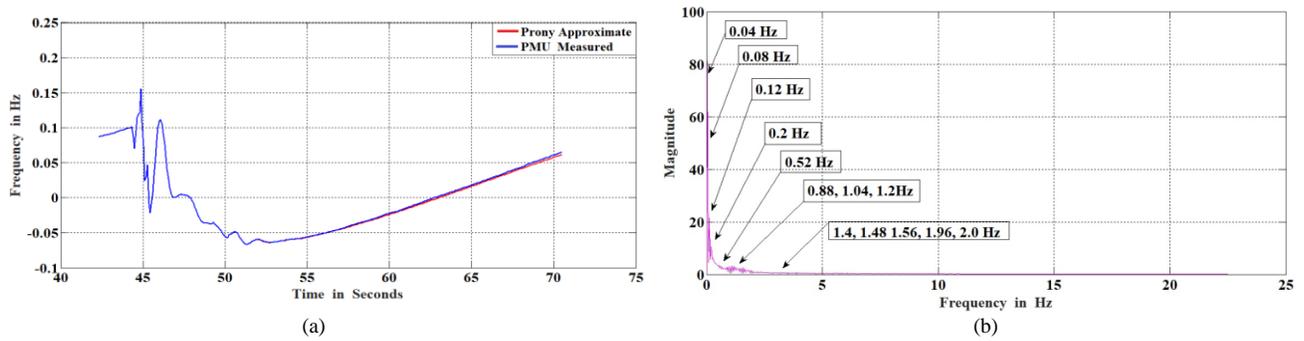

Fig.9. PMU data analysis plots for 400kV Dehgam substation (a) plot for Prony approximate and PMU measured frequency response, and (b) FFT Plot indicating dominant frequency modes and related magnitude for frequency.

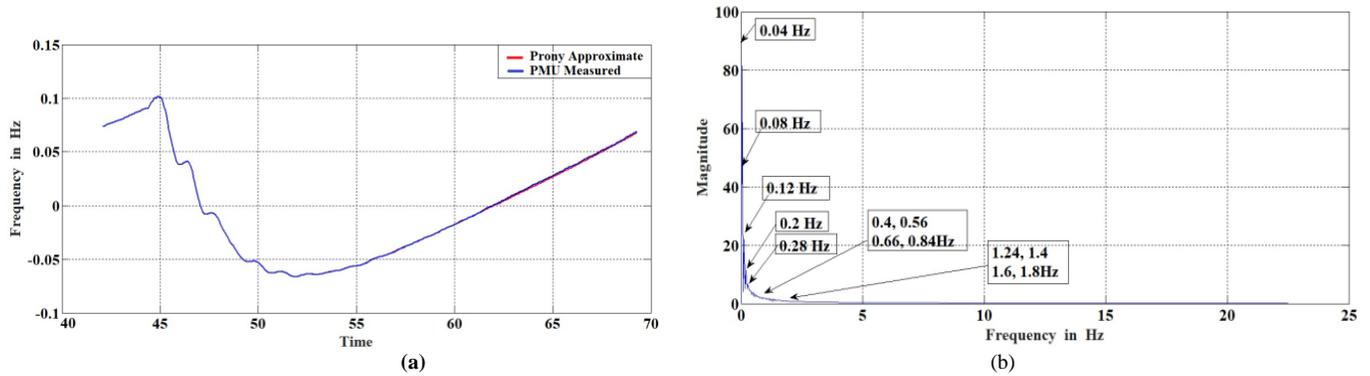

Fig.10. PMU data analysis plots for 400kV Korba substation (a) plot for Prony approximate and PMU measured frequency response, and (b) FFT Plot indicating dominant frequency modes and related magnitude for frequency.

Similarly, the responses of PMUs in other important states of Western India; Gujarat, Madhya Pradesh, and Chhattisgarh are also considered for Prony and FFT analysis.Fig.8 shows 400kV Asoj substation PMU data analysis plots within the GETCL grid.Fig.8 (a) shows frequency behavior as shown by PMU at 400kV Asoj substation during CGPL-UMPP disturbance. 400kV Asoj is geographically nearer to the CGPL-UMPP compared to other locations in GETCL. Therefore, PMU data from Asoj is taken into consideration for analysis purposes. Fig. 8(b) shows a plot for Prony approximate and PMU measured frequency response for 400kV Asoj substation in the GETCL system. Fig.8(c) indicates a magnified part of the Fourier transform plot obtained for the same. It also highlights the fact that 400kV Asoj PMU nearer to CGPL-UMPP shows oscillations in the grid more distinctly when compared to those shown by PMUs located away from the CGPL-UMPP. Fig.9 shows 400kV Dehgam substation PMU data analysis plots within the GETCL system. Fig.9 (a) shows a plot for Prony approximate and PMU measured frequency response for 400kV Dehgam substation, whereas Fig. 9(b) indicates a magnified part of Fourier transform plot for the same. Fig.10 shows 400kV Korba substation PMU data analysis plots within the CSPTCL grid. Fig. 10(a) indicates a plot for Prony approximate and PMU measured frequency response, whereas Fig. 10(b) shows a magnified part of the Fourier transform plot obtained for the same. Fig.11 shows Vindhyachal Super Thermal Power Station (VSTPS) attached 400kV VSTPS substation PMU data analysis plots. Fig. 11(a) shows a plot for Prony approximate and PMU measured frequency response for 400kV VSTPS substation, whereas Fig.11(b) indicates a magnified part of the Fourier transform plot obtained for the same.

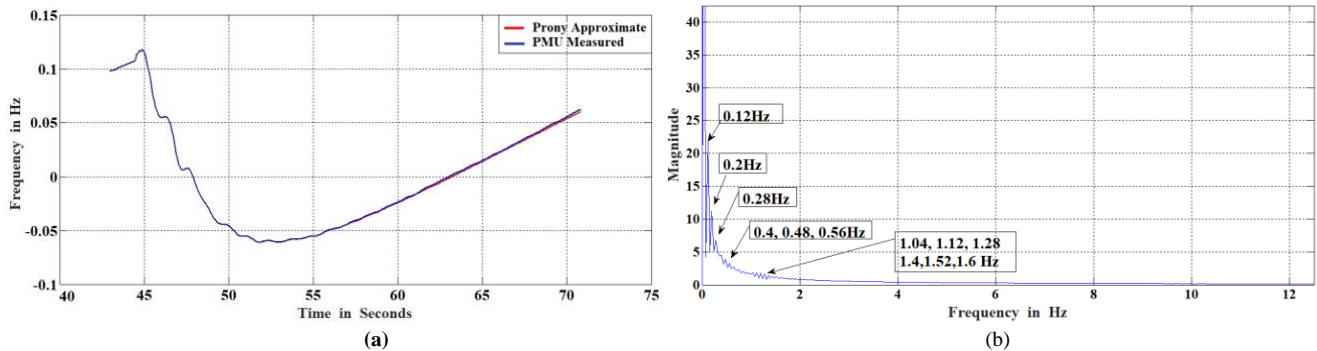

Fig.11. PMU data analysis plots for 400kV VSTPS substation (a) plot for Prony approximate and PMU measured frequency response, and (b) FFT Plot indicating dominant frequency modes and related magnitude for frequency.

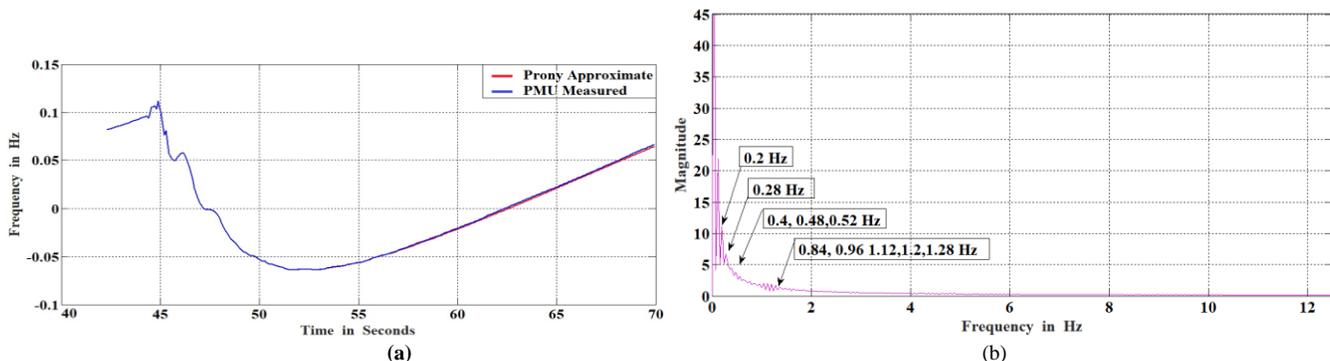

Fig.12. PMU data analysis plots for 400kV Itarasi substation (a) plot for Prony approximate and PMU measured frequency response, and (b) FFT Plot indicating dominant frequency modes and related magnitude for frequency.

Fig.12 shows 400kV Itarasi substation PMU data analysis plots. Fig. 12(a) shows a plot for Prony approximate and PMU measured frequency response, whereas Fig.12 (b) indicates a magnified part of the Fourier transform plot for the same. It indicates the presence of modal frequencies in the range of 0.61-0.68Hz, 0.81-0.87Hz, 0.51Hz, and 0.52 Hz prominently. The results also exhibit negative damping in the measurements of PMUs from western India considered for analysis of CGPL-UMPP occurrence. It indicates the presence of growing oscillations. The summary of the Prony analysis illustrating amplitude, damping, and frequency in Hz for CGPL-UMPP occurrence is tabulated in Table 1. It shows prominent modal frequencies with negative damping. This analysis is based on the PMU data received from different substations in the WR grid.

Table 1. Summary of Prony Analysis for CGPL-UMPP event.

| 220 kV Eklahre Substation (MSETCL) | | | | 400 kV Dehgam Substation (GETCL) | | | |
|---|---|---|---|---|---|---|---|
| Sr. No. | Amplitude | Damping | Frequency in Hz | Sr. No. | Amplitude | Damping | Frequency in Hz |
| 1 | 0.16 | -0.58 | 0.67 | 6 | 0.042 | -0.36 | 0.51 |
| 2 | 0.11 | -0.81 | 0.84 | 7 | 0.041 | -0.47 | 0.30 |
| 3 | 0.046 | -0.47 | 0.49 | **400 kV Korba Substation (CSPTCL)** | | | |
| 4 | 0.032 | -0.58 | 0.2 | 1 | 0.13 | -0.69 | 0.7 |
| 5 | 0.02 | -1.3 | 1.4 | 2 | 0.066 | -0.58 | 0.25 |
| **400 kV Chandrapur Substation (MSETCL)** | | | | 3 | 0.065 | -0.35 | 0.63 |
| 1 | 0.19 | -1.1 | 0.57 | 4 | 0.057 | -0.42 | 0.36 |
| 2 | 0.09 | -0.55 | 0.6 | 5 | 0.026 | -0.27 | 0.85 |
| 3 | 0.032 | -0.39 | 0.84 | 6 | 0.01 | -0.24 | 0.52 |
| 4 | 0.029 | -0.5 | 0.18 | 7 | 0.033 | -0.61 | 0.94 |
| 5 | 0.013 | -0.44 | 0.31 | **400 kV VSTPS Substation (MPPTCL)** | | | |
| **400 kV Asoj Substation (GETCL)** | | | | 1 | 0.093 | -0.69 | 0.15 |
| 1 | 0.69 | -0.93 | 0.72 | 2 | 0.034 | -1.2 | 0.97 |
| 2 | 0.44 | -0.56 | 0.68 | 3 | 0.034 | -0.62 | 0.71 |
| 3 | 0.25 | -0.69 | 0.47 | 4 | 0.014 | -0.25 | 0.85 |
| 4 | 0.12 | -0.48 | 0.85 | 5 | 0.011 | -0.27 | 0.52 |
| **400 kV Dehgam Substation (GETCL)** | | | | **400 kV Itarasi Substation (MPPTCL)** | | | |
| 1 | 0.29 | -0.48 | 0.67 | 1 | 0.064 | -0.57 | 0.71 |
| 2 | 0.19 | -0.50 | 0.87 | 2 | 0.043 | -0.47 | 0.61 |
| 3 | 0.13 | -0.69 | 1.1 | 3 | 0.022 | -0.49 | 0.19 |
| 4 | 0.077 | -0.85 | 1.6 | 4 | 0.016 | -0.38 | 0.37 |
| 5 | 0.061 | -0.46 | 0.78 | 5 | 0.015 | -0.30 | 0.52 |

The proposed algorithm carried out Prony and FFT analysis of real-time PMU data as explained above. It highlights prominently two types of distinct oscillatory modes observed in the WR grid of India during CGPL-UMPP occurrence [33]. They can be categorized as "Inter-area mode" of oscillations in the range of 0.1-0.8 Hz and "Local mode" of oscillations in the range of 0.7-2.0Hz as seen prominently. The local mode oscillatory frequency of 0.84Hz and 0.85Hz is widely shown by PMU measurements throughout the WR grid. The inter-area mode of the frequency of 0.52 Hz is prominently indicated by PMU measurements almost throughout the WR grid. In addition to this, other local plant modes of frequencies were observed during this disturbance are presented above in the results of Prony and FFT analysis. Indian electrical power grid during various events in the past has exhibited oscillation frequencies in the range of 0.5-0.55Hz as characteristic oscillations between the Western and Eastern regional grid of India. These inter-regional area modes of oscillations in the Indian context were also seen getting excited during the CGPL-UMPP blackout in Gujarat.

### 4.2 Case Study-2: HVDC Spikes in the MSETCL grid

MSETCL is the only State Transmission Utility (STU) in India that owns and operates a ± 500kV Bipolar HVDC link between Chandrapur in Eastern and Padghe in the Western part of Maharashtra. The capacity of this link is 1500 MW. On 7[th] October 2015, pole-1 was under maintenance since 7:56 hrs. Therefore, Pole-2 was functioning in-ground return (Monopolar) mode. Thus, the full return current flows through the electrode line connected to the earth electrode station at Anjur near the Padghe substation [25]. The electrode line used for ground return comprises two parallel conductors EL-1 and EL-2 emanating at the Padghe substation and terminating at the Anjur Electrode station as shown in Fig.13 (a). Electrode unbalances supervision alarm designated by the HVDC control feature got triggered when an uneven current distributed between conductors EL-1 and EL-2. In an investigation, it was found that one of the conductor; EL-1 was charred partly and it was hanging on a few strands. This conductor was partly broken. Thus, a high impedance path was formed that gave uneven sharing of current between EL-1 and EL-2. This is the root cause of the alarm. It was seen that alarms for the same kept on toggling after every two minutes by a timer. This was indicated by spikes in the system parameters of PMU measurements. A snapshot of spikes in frequency observed every two minutes on the synchrophasor system in MSETCL during this event is depicted in Fig. 13(b).

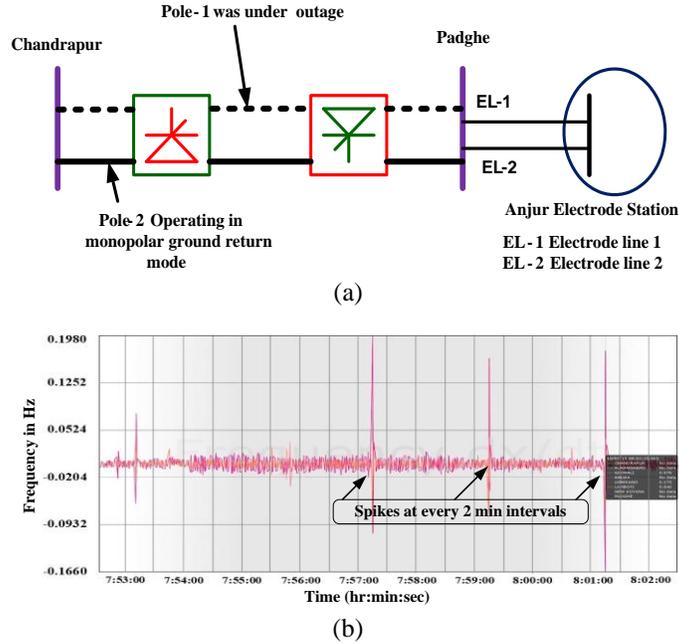

Fig.13. HVDC event (a) Chandrapur-Padghe HVDC bipolar line and electrode feeders at Anjur, and (b) spikes during HVDC event in MSETCL system.

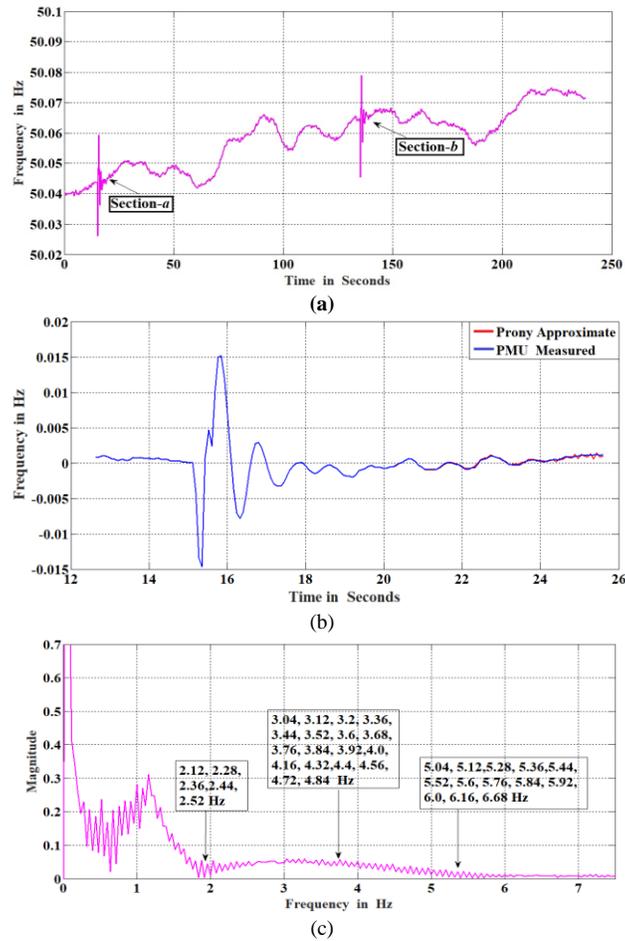

Fig.14. PMU data analysis plots for 220 kV Boisar substation (a) specific data range of PMU measured frequency considered for analysis, (b) plot for Prony approximate and PMU measured frequency response, and (c) FFT Plot indicating dominant frequency modes and related magnitude for frequency.

Table 2. Summary of Prony Analysis for HVDC event.

| 220kV Boisar Substation | | | |
|---|---|---|---|
| Sr. No. | Amplitude | Damping | Frequency in Hz |
| 1 | 0.31 | -0.21 | 4.0 |
| 2 | 0.12 | -0.69 | 4.2 |
| 3 | 0.03 | -0.39 | 3.5 |
| 4 | 0.026 | -0.37 | 4.1 |
| 5 | 0.016 | -0.33 | 3.8 |
| 7 | 0.015 | -0.32 | 3.3 |
| 8 | 0.014 | -0.33 | 3.2 |
| 9 | 0.012 | -0.38 | 4.3 |
| 10 | 0.001 | -0.30 | 3.7 |

It also shows a time interval of two-minute for such spikes. Fig.14 shows 220kV Boisar substation PMU data analysis plots within the MSETCL grid.Fig.14 (a) shows the frequency of PMU at 220 kV Boisar substation in WR grid considered for analysis purpose using proposed algorithm. In this, section 'a' of a signal indicates the first spike, and section 'b' indicates another spike as observed during this event. This phenomenon is analyzed to understand the kind of oscillations present in the system during this incidence.Fig.14 (b) shows a plot for Prony approximate and PMU Measured frequency response of 220 kV Boisar substation for section 'a'. Table 2 depicts the results of Prony analysis based on PMU data from 220kV Boisar substations.Fig.14(c) shows the magnified part of the Fourier transform plot obtained for 220kV Boisar indicating control modes of oscillatory frequency along with respective magnitude. It is observed that oscillation modes in the range of 3.2-6.68Hz were observed prominently with negative damping during this incidence. These control modes of oscillations are related to the "Electrode Unbalance Supervision" feature of HVDC initiated during this event. This anomaly of spikes was arrested only after ramping down power carried by Pole-2 from 750 MW to 600MW around 8.50 AM. These spikes along with respective control modes of oscillations as observed during this event were not seen before and after this incidence, clearly correlate them with the event described above [33].

This section presented two distinct case studies based on the application of the proposed algorithm based on Prony and FFT methods used to process PMU data in the WR grid of India. This gives better insight into the oscillatory behavior by analyzing inter-area mode and local mode of oscillations in case study 1. In case study 2, control modes of oscillations are observed. This knowledge regarding oscillations in the grid shall assist the system planner and operator in mitigating them with required actions. This can be achieved by the system operator in real-time, by adopting measures like generation scheduling, load shedding, activating the Special Protection Schemes (SPS), controlling line flows, utilizing HVDC or flexible AC transmission systems (FACTS) controllers to damp the oscillations for stable and reliable system operation

## 5. Conclusions

This paper focuses on the application of synchrophasor technology for situational awareness in a power system. The various modes of oscillations are described. An application algorithm is recommended for the detection of low-frequency detection. This algorithm is based on Prony and FFT methods. The high-frequency noise is eliminated by using empirical mode detection band pass filter. Two case studies in respect of westerns regional Indian power grid are demonstrated. The first case study of analyzing CGPL-UMPP generation complex related blackout. It highlights the dominant presence of "Inter-area mode of oscillations" and "Local mode of oscillations". It is observed that how oscillations have propagated from one part of the to another part of the power system during this disturbance. The second case study related to HVDC system operations correlates to "Control modes" of oscillations with events observed.

This information regarding the types of oscillations gives awareness about the power grid situation. This situational awareness shall assist the system operator to initiate proper steps for stable and reliable system operation and to maintain grid discipline. It can be accomplished by mitigating oscillations with required actions with the State and Regional Load Dispatch Centers. By adopting corrective measures in real-time like load shedding, generation scheduling, application of special protection schemes (SPS), controlling load flows through transmission lines, effective utilization of HVDC and/or flexible AC transmission systems (FACTS) devices..